\documentclass[10pt,twocolumn,letterpaper]{article}

\usepackage{cvpr}              

\usepackage{url}
\usepackage{multirow}
\usepackage[ruled,vlined]{algorithm2e}
\usepackage{amsmath,bm}
\usepackage{amsfonts}
\usepackage{amssymb}
\usepackage{amsthm}

\usepackage{wrapfig,lipsum,booktabs}

\renewcommand\footnotemark{}

\usepackage[dvipsnames]{xcolor}

\usepackage{graphicx}
\usepackage{amsmath}
\usepackage{amssymb}
\usepackage{multirow}
\usepackage{booktabs}

\usepackage{wrapfig}

\definecolor{cvprblue}{rgb}{0.21,0.49,0.74}
\usepackage[pagebackref,breaklinks,colorlinks,citecolor=cvprblue]{hyperref}

\def\confName{CVPR}
\def\confName{CVPR}
\def\confYear{2024}

\title{LAGA: Layered 3D Avatar Generation and Customization via Gaussian Splatting}

\author{
Jia Gong\textsuperscript{1,2\dag}
~~~ Shengyu Ji\textsuperscript{2\dag}
~~~ Lin Geng Foo\textsuperscript{1}
~~~ Kang Chen\textsuperscript{2}
~~~ Hossein Rahmani\textsuperscript{3} 
~~~ Jun Liu\textsuperscript{1\ddag} \\
\textsuperscript{1}Singapore University of Technology and Design ~~
\textsuperscript{2}Netease ~~
\textsuperscript{3}Lancaster University\\
{\tt\small jia\_gong@mymail.sutd.edu.sg, jishengyu@corp.netease.com, lingeng\_foo@mymail.sutd.edu.sg,}\\ 
{\tt\small ckn6763@corp.netease.com, h.rahmani@lancaster.ac.uk, jun\_liu@sutd.edu.sg} 
}

\newcommand{\nickname}{LAGA}

\usepackage{tikz}

\newcommand*\circled[1]{\tikz[baseline=(char.base)]{\node[shape=circle,draw,inner sep=0.5pt] (char) {#1};}}

\newcommand\blfootnote[1]{%
  \begingroup
  \renewcommand\thefootnote{}\footnote{#1}%
  \addtocounter{footnote}{-1}%
  \endgroup
}

\begin{document}

\twocolumn[{
\renewcommand\twocolumn[1][]{#1}
\maketitle
\pagestyle{empty}
\thispagestyle{empty}
\begin{center}
\centering
\captionsetup{type=figure}
\includegraphics[width=\textwidth]{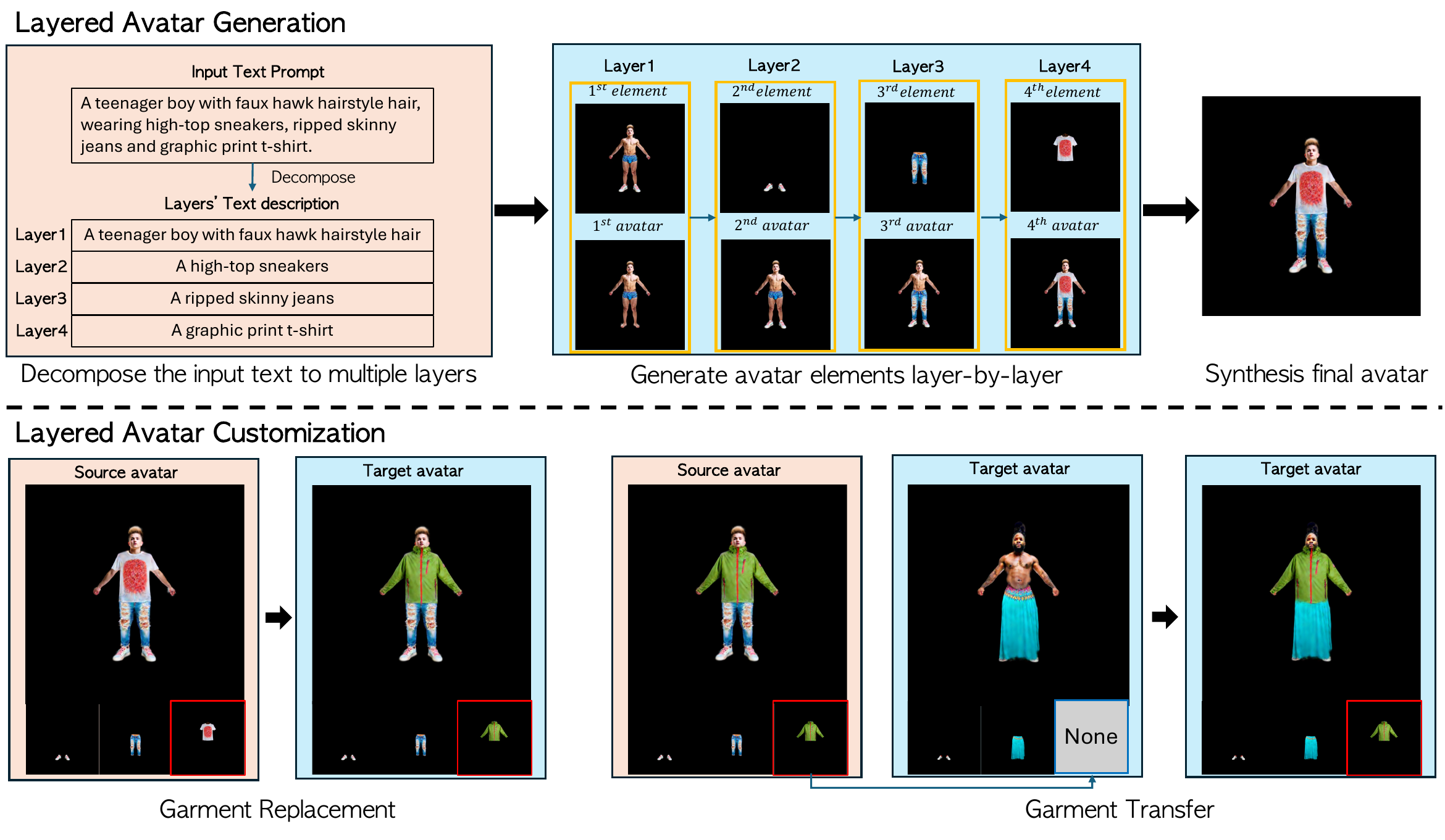}
\captionof{figure}{
We present LAGA, a novel layered avatar generation framework based on Gaussian Splatting (GS). With the layered structure, our generated clothed avatar can be decomposed to a human body with multiple individual garments, allowing users to assemble and edit specific garments to create new variations.
}
\label{fig:teaser}
\end{center}
}]

\blfootnote{\dag~Equal contribution; 
 \ddag~Corresponding author}
 
\begin{abstract}
Creating and customizing a 3D clothed avatar from textual descriptions is a critical and challenging task. Traditional methods often treat the human body and clothing as inseparable, limiting users' ability to freely mix and match garments. In response to this limitation, we present LAyered Gaussian Avatar ({\nickname}), a carefully designed framework enabling the creation of high-fidelity decomposable avatars with diverse garments.
By decoupling garments from avatar, our framework empowers users to conviniently edit avatars at the garment level. Our approach begins by modeling the avatar using a set of Gaussian points organized in a layered structure, where each layer corresponds to a specific garment or the human body itself. To generate high-quality garments for each layer, we introduce a coarse-to-fine strategy for diverse garment generation and a novel dual-SDS loss function to maintain coherence between the generated garments and avatar components, including the human body and other garments. Moreover, we introduce three regularization losses to guide the movement of Gaussians for garment transfer, allowing garments to be freely transferred to various avatars.
Extensive experimentation demonstrates that our approach surpasses existing methods in the generation of 3D clothed humans.
Project page: 
https://gongjia0208.github.io/LAGA/
\end{abstract}

\section{Introduction}
The generation of 3D avatars is an important task that holds immense significance across various industries, including film, gaming, and fashion. 
However, traditional methods for 3D avatar generation often rely on skilled engineers employing specialized software tools \cite{favalli2012multiview} or require the usage of scanners to scan specific actors \cite{feng2017just}, demanding considerable human effort and resources.
Benefited by the developments in generative models \cite{foo2023ai, peng2024harnessing, zhang2024diff, editing}, several research works have attempted to simplify the 3D avatar generation process through large-scale 3D generative models \cite{hong2023lrm,wang2023rodin} or leveraging robust 2D text-to-image priors to generate 3D humans from text prompts \cite{poole2022dreamfusion,cao2023dreamavatar,huang2024dreamwaltz}.
However, despite the significant progress, most works still treat the avatar as a singular entity, lacking the capability to separate garments from the avatar itself.
This inherent limitation presents challenges in avatar customization, particularly in scenarios where users want to decorate diverse clothing and accessories for specific characters, such as in gaming or virtual reality environments.

To address this challenge, a promising approach is to create a \textit{decomposable avatar} {where the garments are separated from the human body}. 
Specifically, a straightforward way is to treat the human body and its garments as separate meshes to generate a disentangled avatar \cite{wang2023disentangled, corona2021smplicit}. 
However, this approach not only requires additional human effort to design garment mesh templates but also encounter difficulties accommodating diverse clothing types due to the inherent geometric constraints of meshes.
In response to this challenge, recent works \cite{feng2022capturing, wang2023humancoser, hu2023humanliff} have explored modeling clothing using Neural Radiance Fields (NeRF) \cite{mildenhall2021nerf}, {which provide better fidelity and flexibility in representing various clothing types}. 
{Yet, due to their implicit representation, NeRF-based approaches} tend to struggle with complex {and} inefficient rendering procedures, requiring multiple network forward passes and/or complex calculations per pixel.
{Besides, NeRF-based approaches also present challenges for applying deformations, making it difficult to transfer the garment when the shape of the human body changes significantly \cite{wang2023humancoser}.}

Recently, 3D Gaussian Splatting (GS) \cite{kerbl20233d} has provided a fresh perspective on 3D asset generation. 
This approach leverages 3D Gaussian points characterized by color, opacity, and density parameters to represent 3D scenes. 
In particular, we observe that the inherent flexibility of their point-cloud-like representation makes GS suitable for generating diverse garments. Meanwhile, the explicit nature of GS grants direct control over the Gaussians, facilitating the customization of garments to suit different body {shapes}.
Building upon these insights, we introduce the LAyered Gaussian Avatar (\nickname) framework to overcome the aforementioned challenges. 
Our framework enables the generation of high-quality 3D avatars with diverse garments, including both tight-fitting and loose clothings, while also allowing for effortless adaptation of garments to different human shapes.
{Specifically, our approach treats a clothed avatar to be comprising multiple layers, with each layer corresponding to a specific component, such as the base avatar, garments, or accessories.}
{To control the location and scale of each component, we create the stack of layers by progressively expanding the SMPL mesh \cite{loper2023smpl} layer-by-layer, initializing Gaussian points based on the expanded mesh and related joints in each layer.}
{Then, we can employ score distillation sampling (SDS) to optimize the Gaussian points at each layer, tapping into the rich 2D knowledge in the pre-trained diffusion model for 3D generation.}

However, we find that there still exists three main challenges to achieve effective generation of decomposable avatars with our pipeline: 
\circled{1} \textbf{Diverse Garment Generation:}  Although initializing {Gaussian points} based on the SMPL model provides a {good basic structure} for locating the avatar component's position and scale, {it can potentially restrict the diversity of generated garments, making it unsuitable for generating garments with shapes diverging significantly from the human body.}
\circled{2} \textbf{Coherence of Generated Garments:} Simply optimizing Gaussian points via SDS loss may lead to garments lacking coherence with other avatar components, detracting from the natural appearance of the avatar.
For example, generating a skirt independently can result in its waistline not closely fitting the human avatar and parts of it occluding the avatar's upper garment, leading to a lower-quality clothed avatar when they are combined together.
\circled{3} \textbf{Difficulty of Garment Transfer:} The \textit{dense and unstructured nature} of GS presents a unique challenge in adapting garments to avatars with diverse body shapes. Unlike meshes, which offer well-defined geometry properties for deformation, controlling thousands of Gaussian points for garment transfer is challenging and requires a multifaceted approach.

To address the above challenges, we propose the following designs. 
\circled{1} First, to facilitate \textit{diverse garment generation}, we propose a coarse-to-fine generation strategy coupled with a density guidance loss. Specifically, we divide the garment generation into two stages: a coarse stage to approximate the overall shape of the target garment and a fine stage for high-quality garment generation. Moreover, we introduce a density guidance loss to guide Gaussian points to match well with the garment shape during optimization.
\circled{2} Second, to ensure coherence between the garments in each layer and the rest of the avatar, we introduce a dual-SDS loss. This loss optimizes local garment-only images for high-quality garment generation while ensuring consistency with other avatar parts through a global {rendering} containing all current garments.
\circled{3} Finally, we propose three regularization losses aimed at guiding the movement of Gaussian points for garment transfer: a Human Fitting Loss to encourage the garment to fit the contours of the human body, a Similarity Loss to preserve the overall shape of the garment during adaptation, and a Visibility Loss to prevent the garment from being obscured by the avatar's existing components. Overall, these losses help guide the thousands of Gaussian points to properly adapt to the target avatar.

In summary, our contributions are as follows:
1) We introduce \nickname, a novel decomposable avatar generation framework capable of producing high-quality decomposable avatar with various garments and support easy garment adaptation between various human body shapes.
2) Our method incorporates various meticulously designed modules to facilitate layered avatar generation and garment adaptation, enabling us to achieve superior quality.
3) Through extensive qualitative and quantitative experiments, we validate the efficacy of our approach. Our method consistently outperforms existing methods, generating avatars of exceptional quality. Moreover, the generated avatars demonstrate a remarkable level of consistency with the corresponding input natural languages.
\section{Related Work}
\label{sec:formatting}

\paragraph{Text-guided 3D Asset Generation.}
Recent text-to-3D generation methods can generally be divided into two main categories:
1) Direct 3D Generation Pipelines: These methods optimize models to directly learn the distribution of 3D explicit representations \cite{he2024dresscode, nichol2022point, jun2023shap, wu2016learning} or implicit representations \cite{jun2023shap,zhang20243d, li2023instant3d}. However, due to the high complexity of 3D data, these methods either struggle to generate complex 3D assets or are restricted to specific categories.
2) 2D-to-3D Lifting Pipelines: These methods generate a 3D scene matching the given prompt by leveraging extensive 2D domain knowledge stored in 2D text-to-image generators. Early approaches \cite{jain2022zero, mohammad2022clip} used the image-text retrieval model, CLIP \cite{radford2021learning}, to guide the image-text alignment in each camera view for 3D generation. 
Recently, leveraging the powerful 2D generation ability of diffusion-based text-to-image models \cite{rombach2022high, saharia2022photorealistic}, several 3D generation techniques \cite{lin2023magic3d, tang2023dreamgaussian, zhu2023hifa} employ SDS \cite{poole2022dreamfusion}, which stochastically distills the 2D knowledge from diffusion models, to generate high-quality 3D assets.
While the above methods have achieved remarkable success in 3D generation, adopting them for decomposable avatar generation remains challenging due to the high complexity of the hierarchical avatar structure and the {huge difficulties involved in generating realistic textures}.

\paragraph{Text-guided 3D Human Generation.}
{To facilitate text-to-3D human generation, most works adopt 2D-to-3D lifting pipelines with various dedicated designs to incorporate human priors.}
For instance, AvatarCLIP \cite{hong2022avatarclip} pioneers the integration of a parametric human model (SMPL \cite{loper2023smpl}) with Neus \cite{wang2021neus}, leveraging CLIP \cite{radford2021learning} for supervising the creation of diverse 3D humans.
More recently, various approaches \cite{kolotouros2024dreamhuman, cao2023dreamavatar, zhang2024avatarverse} have adopted score distillation sampling (SDS) for generating high-quality clothed humans.
Specifically, DreamHuman \cite{kolotouros2024dreamhuman} introduces a pose-conditioned NeRF model for animatable 3D clothed human generation. 
Both DreamAvatar \cite{cao2023dreamavatar} and AvatarCraft \cite{jiang2023avatarcraft} utilize the pose and shape parameters of SMPL as a guiding prior for high-quality human synthesis.
Further advancements address specific challenges and enhance realism. DreamWaltz \cite{huang2024dreamwaltz} tackles the Janus (multi-face) problem by implementing an occlusion-aware SDS loss with skeleton-based conditioning techniques. AvatarVerse \cite{zhang2024avatarverse} replaces human skeleton conditions in conditional diffusion models with DensePose maps, enhancing view consistency in 3D human generation. TADA \cite{liao2023tada} replaces the NeRF representation with a deformable SMPL-X mesh and optimizes texture UV-maps for avatar rendering, making the generated avatars more suitable for computer graphics workflows. HumanNorm \cite{huang2023humannorm} refines diffusion models to generate normal maps, enriching the geometric fidelity of the resulting avatars.
Recently, HumanGaussian \cite{liu2023humangaussian} explores modeling avatars via 3D GS, generating high-quality clothed humans with fast rendering speeds. 
However, these methods focus on {generating human models as a single entity}, and thus lack the ability to effectively decouple bodies and clothing.
Moreover, in contrast to \cite{liu2023humangaussian}, which primarily focuses on utilizing GS for better avatar rendering performance, our key contribution lies in recognizing the high flexibility and controllability of GS {due to its explicit nature}, which unlocks significant potential for more flexible, layered avatar generation.

\paragraph{Layered Avatar Modeling.} 
Early methods for modeling layered avatars treat the human body and its garments as two separate meshes to generate disentangled avatars \cite{wang2023disentangled, corona2021smplicit, jiang2020bcnet, xiang2021modeling, zhu2020deep}. 
However, this approach requires additional human effort to design garment mesh templates and faces difficulties accommodating diverse clothing types due to the inherent geometric constraints of meshes.
In response to this challenge, recent works \cite{feng2022capturing, wang2023humancoser, hu2023humanliff} have explored modeling clothing using NeRFs \cite{mildenhall2021nerf}, which provide better fidelity and flexibility in representing various clothing types. Specifically, HumanLiff \cite{hu2023humanliff} generates the avatar in a layer-wise manner, presenting the human with clothing in each layer via a triplane neural feature. However, the features of the human body and garments are still not disentangled, limiting the ability for garment editing. 
Conversely, other existing works \cite{feng2022capturing, wang2023humancoser} model the human body and garments separately, but due to their implicit representation, the garments generated by these methods cannot be easily deformed, making them transferrable only between avatars with similar human shapes \cite{wang2023humancoser}.
In contrast, our method can generate decomposable clothed avatars with diverse, replaceable garments and supports garment transfer between avatars with various human shapes.

\section{Method}
\begin{figure*}[t]
\centering
\includegraphics[width=0.9\linewidth]{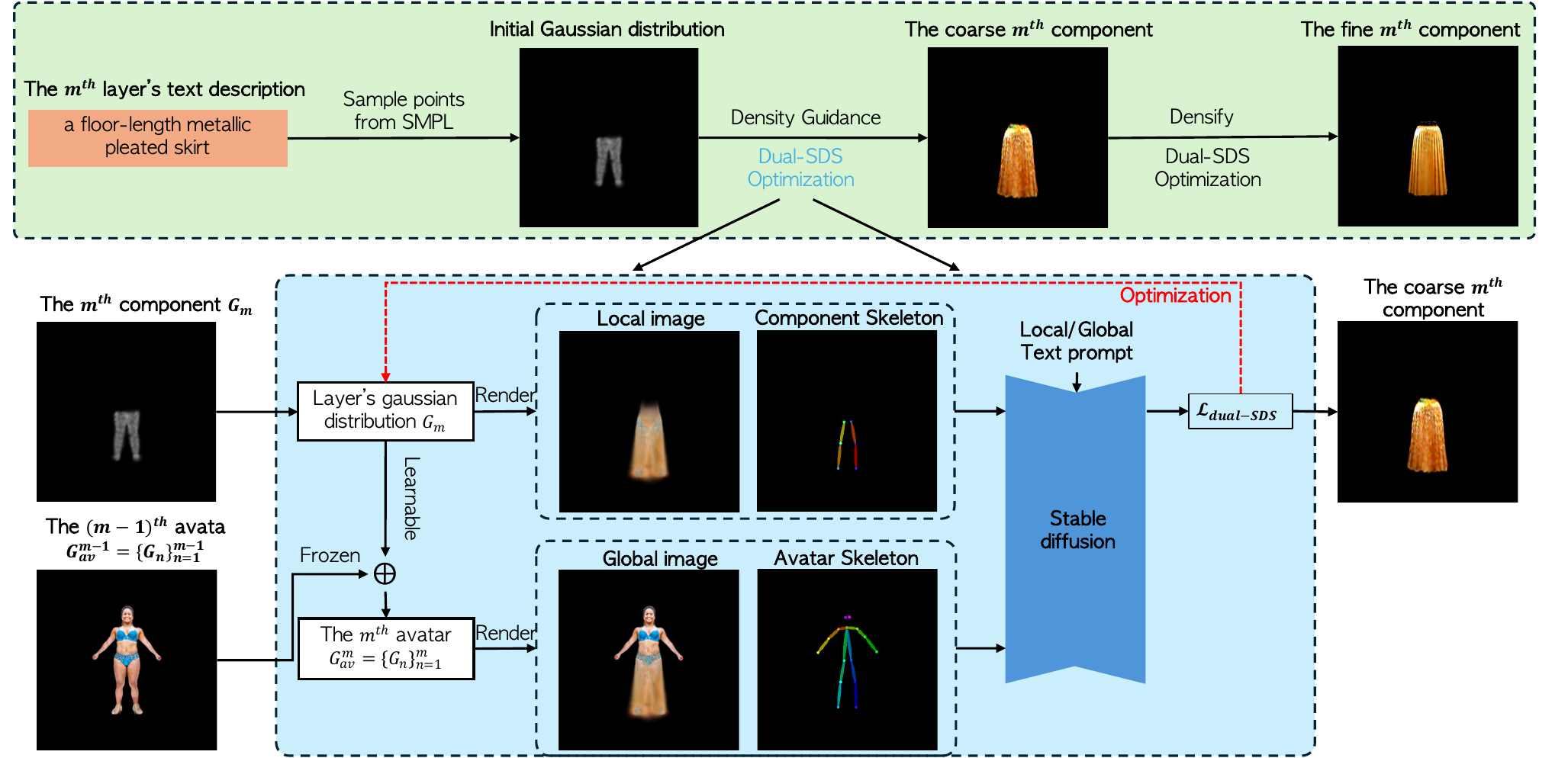}
\caption{\textbf{Overview of the avatar component generation process in each layer}. {As outlined in the green box, our generation process of each layer mainly consists of three steps: (a) sparse initialization of Gaussian points, (b) density guidance to obtain coarse garment, (c) densification to obtain fine garment,}
In the beginning, based on the given layer's text description, we initialize a set of sparse Gaussian points using the parametric human model (SMPL) and associated joints. 
Then, these points are refined to approximate the broad shape of the target component in the coarse stage. 
Subsequently, in the fine stage, we densify the Gaussians to capture finer details and sharper features of the avatar component, aiming for high-quality results. 
To ensure coherence with other generated avatar components, a dual-SDS loss {(as presented in the blue box)} is introduced to optimize the Gaussian points in both coarse and fine stages. This loss function optimizes Gaussians from both local and global perspectives, enhancing the quality and coherence of the generated avatar component.}
\label{fig:pipeline}
\end{figure*}

We present LAyered Gaussian Avatar ({\nickname}), a method for generating decomposable clothed avatars with diverse, interchangeable garments. 
First, to facilitate better understanding, we introduce some important preliminaries regarding SDS and 3D GS in Section \ref{sec:preliminaries}. 
Subsequently, we introduce our method in two parts: how to generate the decomposable avatar (covered in Section \ref{sec:method_generation}), and how to perform garment transfer (covered in Section \ref{sec:method_customization}).
Specifically, in Section \ref{sec:method_generation} we present our avatar generation framework, which includes a coarse-to-fine strategy for diverse garment generation and a dual-SDS loss for coherent garment generation. 
Then, in Section \ref{sec:method_customization}, we introduce three regularization losses to facilitate garment transfer.
Our overall framework is illustrated in Fig. \ref{fig:teaser}.

\subsection{Preliminaries}
\label{sec:preliminaries}

\noindent \textbf{Score Distillation Sampling (SDS)} is introduced in DreamFusion \cite{poole2022dreamfusion} for refining 3D representations by leveraging a 2D pre-trained diffusion generator. Specifically, a 3D scene, parameterized by $\theta$, is optimized to render images that match with the {data distribution of natural images} learned by diffusion model $\phi$ across various noise levels.
In practical implementation, DreamFusion employs a {text-to-image diffusion model \cite{saharia2022photorealistic}} as the score estimator $\bm{\epsilon}_{\phi}(\mathbf{c}_t; y)$, which {predicts} the sampled noise $\bm{\epsilon}_{\phi}$ based on the noisy image $\mathbf{c}_t$, text embedding $y$, and timestep $t$. SDS {optimizes} 3D scenes ($\theta$) through gradient descent with respect to $\theta$ as follows:
\setlength{\abovedisplayskip}{0pt}
\setlength{\belowdisplayskip}{0pt}
\begin{align}
\label{eq:sds}
\small
\nabla_\theta \mathcal{L}_{\mathrm{SDS}}=\mathbb{E}_{\bm{\epsilon}, t}\left[w_t\left(\bm{\epsilon}_\phi\left(\mathbf{c}_t ; y\right)-\bm{\epsilon}\right) \frac{\partial \mathbf{c}}{\partial \theta}\right],
\end{align}
where $\bm{\epsilon}\sim\mathcal{N}(\mathbf{0}, \mathbf{I})$ is Gaussian noise, $\mathbf{c}_t=\alpha_t\mathbf{c}+\sigma_t\bm{\epsilon}$ is the noised image; $\alpha_t$, $\sigma_t$, and $w_t$ are noise {hyperparameters}.

\noindent \textbf{3D Gaussian Splatting (3D GS)} \cite{kerbl20233d} introduces an efficient yet effective approach for 3D scene representation. 
3D GS represents the scene using a collection of anisotropic Gaussians defined by their center position $\mu$, covariance $\Sigma$, color $c$, and opacity $\alpha$. During rendering, a ray $r$ is cast from the center of the camera, and the color and density of the 3D Gaussians that the ray intersects are computed along the ray. The rendering process is as follows:
\begin{align}
\label{eq:gs}
\begin{aligned}
G\left(p, \mu_i, \Sigma_i\right) = \exp (-\frac{1}{2} (p - \mu_i)^\mathsf{T} \Sigma_i^{-1} (p - \mu_i))&, \\
\mathbf{c}(r)=\sum_{i \in \mathcal{M}} c_i \sigma_i \prod_{j=1}^{i-1}\left(1-\sigma_j\right), {\text{where~~}} \sigma_i=\alpha_i G\left(p, \mu_i, \Sigma_i\right)&,
\end{aligned}
\end{align}
where $\mathbf{c}(r)$ is the color value of the pixel in the 2D image $\mathbf{c}$ contributed by the ray $r$; {$p$} is the location of queried point on the ray $r$; $\mu_i$, $\Sigma_i$, $c_i$, $\alpha_i$, and $\sigma_i$ are the center position, covariance, color, opacity, and density of the $i$-th Gaussian respectively; $G\left(p, \mu_i, \Sigma_i\right)$ is the value of the $i$-th Gaussian at point $p$; {$\mathcal{M}$} denotes the set of 3D Gaussians in this tile.

\subsection{Layered Avatar Generation}\label{sec:method_generation}

In this section, we present our proposed approach for generating a decomposable avatar in a layer-by-layer manner. As shown in Fig. \ref{fig:teaser}, for an avatar with $N-1$ garments described in the text prompt, we first create $N$ layers to represent the human body and garments independently. 
Then, we sequentially generate the human body and garments, aiming to optimize the Gaussian points in each layer to produce a component (i.e., human body or garments) that matches its text description and integrates with other existing avatar components seamlessly.
{However, there are two notable challenges:}
\circled{1} \textbf{Diverse Garment Generation:} {During initialization of the 3D avatar,} although initializing {Gaussian points} based on the SMPL model provides a {good basic structure} for locating the avatar component's position and scale, {it can potentially restrict the diversity of generated garments, making it unsuitable for generating garments with shapes diverging significantly from the human body.}
\circled{2} \textbf{Coherence of Generated Garments:} {During optimization,} simply optimizing Gaussian points via SDS loss may lead to garments lacking coherence with existing avatar components, detracting from the natural appearance of the avatar.
To address the above challenges, we propose a \textit{Coarse-to-Fine Generation Strategy} and \textit{Dual-SDS Loss} to tackle challenge \circled{1} and challenge \circled{2} respectively. 
We explain these two designs in detail below.

\paragraph{Coarse-to-Fine Strategy.}
Facilitating diverse garment generation for clothed avatars (Challenge \circled{1}) is challenging because we need to satisfy two requirements: 1) the garment should be suitable for the target avatar; 2) yet, the 3D GS Gaussian points need to be optimized towards a diverse range of garments.
Notably, it is challenging to simultaneously achieve both requirements.
For instance, an intuitive approach to the first problem is initializing the Gaussian points by sampling points from the SMPL-X mesh, which provides a robust foundation for determining the position and scale of the garment. 
However, this approach makes the second problem harder to solve, as the shapes of many loose garments (e.g., skirt) differ significantly from the human body, and 
an inappropriate initialization of GS will lead to a significant performance drop \cite{tang2023dreamgaussian,kerbl20233d}, as shown in Fig. \ref{fig:ab1}. 
Therefore, to address these challenges and achieve diverse garment generation, we divide the garment generation process into two stages as shown in Fig. \ref{fig:pipeline}, which we call \textit{Coarse-to-Fine Strategy}.
Specifically, in the \textbf{Coarse Stage}, we initialize a sparse set of Gaussian points and optimize {them} to approximate the overall shape of the target garment, allowing the Gaussian points to be initialized in diverse shapes accordingly.
Then, in the \textbf{Fine Stage}, to capture sharper and more detailed garment features, we densify the Gaussian points, allowing them to be more suitable for the target avatar.

In the \textbf{Coarse Stage}, we begin by initializing the Gaussian points {at each layer by} sampling a small number of points (5,000 points) from the SMPL-X mesh. To focus these points on the target garment {at each layer}, we query relevant human joints to generate a 3D bounding box and remove Gaussian points outside this box. 
{By performing this initialization at each layer, we can obtain the set of sparse Gaussian points at each $m$-th layer, which we denote as $\mathbf{G}_m$.}

Next, we aim to optimize $\mathbf{G}_m$ to approximate the coarse shape of the {$m$-th avatar component} described in the {text prompt}. 
A straightforward approach is to adopt the SDS loss {(discussed in Section \ref{sec:preliminaries})} to optimize \(\mathbf{G}_m\) to match the target garment.
However, SDS loss primarily focuses on optimizing \(\mathbf{G}_m\) to produce a natural-looking 2D image in each view independently, which is stochastic \cite{song2020score} and lacks strong geometry constraints. 
There is no explicit regularization process to control the density of Gaussian points throughout the avatar during SDS optimization, and thus to model the avatar, SDS can often optimize the Gaussian points in certain areas to be sparser but larger, especially for the areas where the initialized Gaussian points were already sparse.
However, this can be sub-optimal, since the Gaussian points may turn out to be overly sparse at some areas, which poses issues with modeling the coarse approximate shape of the component (see Fig. \ref{fig:ab1} for visualization).

Therefore, to encourage the Gaussian points to be spread evenly for a better coarse approximation of the component's shape,
we propose incorporating density guidance into optimization to ensure the Gaussian points are more evenly distributed to address this issue.
Specifically, we regard the opacity of each Gaussian point as its density in the 3D space and then render a 2D opacity map of GS to {represent} the density distribution of $G_m$.
Formally, similar to Eq. \ref{eq:gs}, the opacity map of \(\mathbf{G}_m\) is computed by accumulating the opacity values along the ray $r$, as shown below:
\setlength{\abovedisplayskip}{0pt}
\setlength{\belowdisplayskip}{0pt}
\begin{align}
\label{eq:gs-opacity}
\mathbf{\alpha}_m(r)=\sum_{i \in \mathcal{M}} \sigma_i \prod_{j=1}^{i-1}\left(1-\sigma_j\right), \text{where}~ \sigma_i=\alpha_i G\left(p, \mu_i, \Sigma_i\right)&,
\end{align}
where $\mathbf{\alpha}_m(r)$ is the value of the 2D opacity map $\mathbf{\alpha}_m$ contributed by the ray $r$; $r$ is a ray cast from the center of the camera; $p$ is the location of queried point on the ray; $\alpha _i$ is the opacity of $i$-th Gaussian and $G\left(p, \mu_i, \Sigma_i\right)$ is the value of the $i$-th Gaussian at the queried point $p$ as defined in Eq.~\ref{eq:gs}.

After that, we capture the areas occupied by the component by creating a binary component mask \(M_m\) via a segmentor \cite{kirillov2023segment} and then optimize the density of Gaussians in these areas to be uniform as shown below:
\setlength{\abovedisplayskip}{0pt}
\setlength{\belowdisplayskip}{0pt}
\begin{align}
\label{eq:loss_density}
\begin{aligned}
\mathcal{L}_{d} = ||M_m-f_n(M_m*\mathbf{\alpha}_m)||^2_2
\end{aligned}
\end{align}
where $f_n$ is a normalization operation that adjusts the values of the masked opacity map $(M_m*\mathbf{\alpha}_m)$ to range between 0 and 1. 
With this strategy, we can effectively control the sparse Gaussian points $G_m$ to fit the coarse shape of the target garment, making it suitable for generating diverse garments, {including garments with shapes that are very different from the human body}.

In the \textbf{Fine Stage}, our goal is to refine \(\mathbf{G}_m\) to {obtain} sharper and more detailed garment features. 
To achieve this, we recurrently upsample the Gaussian points and then optimize them to generate the high-quality avatar component via SDS loss.
During each upsampling step, instead of simply duplicating \(\mathbf{G}_m\) to create a denser distribution, we propose to duplicate the Gaussian points \(\mathbf{G}_m\) with several perturbations to better capture the detailed variations in the local object area. 
Specifically, we duplicate the existing Gaussian points \(\mathbf{G}_m\) to create {another set of Gaussian points}, denoted as \(G_d\), and then {perturb their} positions and colors as follows:
\begin{equation}
\label{eq:noise_sampling}
\bm{\mu}_{d}' = \bm{\mu}_d + \bm{\epsilon}_{d},\;\; \bm{c}_{d}' = \bm{c}_d + \bm{\epsilon}_c,
\end{equation}
where $\bm{\mu}_{d}$ and $\bm{c}_{d}$ {are the original positions and colors of the Gaussian points in ${G}_d$}, $\bm{\mu}_{d}'$ and $\bm{c}_{d}'$ are the updated positions and colors, $\bm{\epsilon}_{d}$ is a small position noise sampled between -0.0005 and 0.0005, and $\bm{\epsilon}_{c}$ is the color noise sampled between 0 and 0.05.
Then, we obtain the denser set of Gaussian points by merging $\mathbf{G}_d$ into $\mathbf{G}_m$ and optimize the updated $\mathbf{G}_m$ via the SDS loss (see Section \ref{sec:preliminaries}) to generate a high-quality garment.

\paragraph{Dual-SDS Loss.}

While {our Coarse-to-Fine strategy above} offers a good framework for controlling the position and scale of each avatar component, optimizing {each component} individually can sometimes lead to a lack of coherence with other parts, resulting in an unnatural appearance. This issue {(Challenge \circled{2})} arises from the shape changes of each component during optimization and the inherent geometric complexity of overlapping areas. For example, optimizing pants from the standard SMPL-X model might not yield a suitable fit for a slender woman. Similarly, independently creating a loose shirt and jeans can lead to issues with occlusion at the waist area, where the shirt and jeans {may} overlap. 
These discrepancies can accumulate and become noticeable, causing the avatar to appear disjointed or proportionally incorrect.

Motivated by the idea that garments not only exist individually but also seamlessly blend into the avatar's overall look, we propose a dual-SDS loss, which optimizes the layer's Gaussian points $G_m$ {while} considering both local and global aspects. 
{At the local level, we focus on optimizing the individual garment by optimizing its images (rendered from $G_m$) to precisely align with the layer's textual description.}
{At the same time, we also consider the global perspective by optimizing the unified image that also incorporates the inner $m-1$ avatar components, up to the $m$-th layer.}
By utilizing the Gaussian points from the layers up to the $m$-th layer ($\{\mathbf{G}_{j}\}^{m}_{j=1}$), this {global} view enables us to optimize $G_m$ to be aware of the overall appearance, resulting in a seamless and natural visual coherence using SDS loss.

{More precisely,} to achieve this, we first combine $\mathbf{G}_m$ with the inner $m-1$ layers ($\{\mathbf{G}_{j}\}^{m-1}_{j=1}$) to obtain the ``global'' avatar for the $m$-th layer as: $\mathbf{G}_{av}^m = \{\mathbf{G}_{j}\}^{m}_{j=1}$.
Then we follow Eq. \ref{eq:gs}  to render a local image $\mathbf{c}^l$ from  $\mathbf{G}_{av}^m$ using the following formulation:
\begin{equation} \label{eq:gs_composition}
\begin{split}
\mathbf{c}^l(r)=\sum_{i \in \mathcal{M}(\mathbf{G}_m,r)} c_i \sigma_i \prod_{j=1}^{i-1}\left(1-\sigma_j\right), \sigma_i=\alpha_i G\left(p, \mu_i, \Sigma_i\right)&.
\end{split}
\end{equation}
where $\mathcal{M} (\mathbf{G}_m, r)$ refers to the set of Gaussian points in $\mathbf{G}_m$ that are along the ray $r$.
Meanwhile, to render a global image $\mathbf{c}^g$, we modify Eq. \ref{eq:gs_composition} by replacing $\mathbf{G}_m$ with $\mathbf{G}_{av}^m$.

{Next, to optimize the avatar components, we apply the SDS loss to the rendered local images $\mathbf{c}^l$ and global images $\mathbf{c}^g$ to encourage them to match the natural images learned by the 2D diffusion generator.}
For our SDS loss, we {follow previous works \cite{huang2024dreamwaltz} to} adopt a 2D human skeleton conditioned diffusion model \cite{zhang2023adding} to enhance multi-view consistency of our human avatar.
Formally, conditioned on the 2D human skeleton $\mathbf{s}$, our dual-SDS loss (modified from Eq. \ref{eq:sds}) for the $m$-th layer is expressed as:
\begin{align}
\label{eq:dual-sds}
\begin{aligned}
\nabla_\theta \mathcal{L}_{\mathrm{Dual-SDS}} &= \lambda_l \cdot \mathbb{E}_{\bm{\epsilon}_{\mathbf{x}^l}, t}\left[w_t\left(\bm{\epsilon}_\phi\left(\mathbf{x}^l_t ; \mathbf{s}, y^l\right)-\bm{\epsilon}_{\mathbf{x}^l}\right) \frac{\partial \mathbf{x}^l}{\partial \theta}\right] \\
&+ \lambda_g \cdot \mathbb{E}_{\bm{\epsilon}_{\mathbf{x}^g}, t}\left[w_t\left(\bm{\epsilon}_\phi\left(\mathbf{x}^g_t ; \mathbf{s}, y^g\right)-\bm{\epsilon}_{\mathbf{x}^g}\right) \frac{\partial \mathbf{x}^g}{\partial \theta}\right],\\
\end{aligned}
\end{align}
where $y^l$ is text prompt of the $m^{th}$ garment; $y^g$ denotes the text prompt of the $m^{th}$ avatar, which is a combination of the human body description and the layer's text description; {$\theta$ represents the parameters of the Gaussian points in the $m$-th layer ($G_m$); and $\lambda_l, \lambda_g$ are two pre-defined hyperparameters}.
To ensure coherent avatar component generation, we replace the SDS loss in the both coarse and fine generation process with our dual-SDS loss (see Figure \ref{fig:pipeline}). 
{Note that, since the bare human body serves as the fundamental avatar component, we solely employ the SDS loss during the human body generation in the first layer (i.e., when $m=1$).}

Overall, by dividing the avatar component generation process into coarse and fine stages, we can optimize sparse Gaussian points to approximate the basic shapes of {diverse} garments and then densify these Gaussians for high-quality garment generation. Additionally, by applying the dual-SDS loss to optimize Gaussians from both local and global perspectives, we ensure coherence between the generated garment and {other} avatar components.

\subsection{Garment Transfer}\label{sec:method_customization}

With the layered structure described in the previous subsections, our avatar can be conveniently divided into multiple components, allowing users the freedom to decorate it as they wish, such as replacing an old garment with a new one, as shown in Fig. \ref{fig:teaser}.
This flexibility sparks an intriguing possibility: could we replace our avatar's garments by transferring garments from other avatars rather than generating entirely new ones?
Note that, although previous methods have attempted this \cite{feng2022capturing, wang2023humancoser}, they are constrained to transferring clothes between avatars with similar body shapes. 
Leveraging the explicit representation of 3D GS and the control it offers, we aim to overcome this limitation by enabling the transfer of garments {between avatars with differing body shapes}.

However, the dense and unstructured nature of GS poses a unique obstacle in adapting garments to avatars with varying body shapes. Unlike meshes, which offer well-defined geometric properties conducive to deformation, controlling thousands of Gaussian points for garment transfer demands a nuanced approach {(Challenge \circled{3})}. 
Here, {to transfer the garment to avatars with a different body shape}, we freeze all parameters of Gaussian points except the position and scale, and introduce three regularization losses to guide the movement of Gaussian points {for adaptation}.

Firstly, {since well-fitting garments (either loose or tight) need to be tailored} to follow the body's natural curves and proportions \cite{gill2015review}, we introduce a Human Fitting Loss $\mathcal{L}_{HF}$ to regularize the shape of the garment $\mathbf{G}_m$. 
This loss function projects the garment and the human body separately onto 2D images, and optimizes the depth map of the garment to match the depth map of the human body in the overlapping areas, encouraging the garment to closely fit to the human contour.
Formally, it can written as: 
\begin{align}
\label{eq:loss_hf}
\mathcal{L}_{HF} = ||\mathbf{d}_{av} - \mathbf{d}_m||^2_2*M_{oc}, ~\text{where}~ M_{oc} = M_{av} \cap M_m
\end{align}
where $\mathbf{d}_{av}$ and $\mathbf{d}_m$ represent the depth map rendered by the target and $\mathbf{G}_m$ respectively, and $M_{oc}$ in a mask that reflects the overlapping area between the garment mask $M_m$ and the target avatar mask $M_{av}$, generated by the segmentor (SAM \cite{kirillov2023segment}).

On the other hand, preserving the overall shape of the garment is crucial for successful transfer. To achieve this, we introduce a Similarity Loss $\mathcal{L}_{ssim}$ that regularizes the transferred garment to resemble its pre-transfer form as:
\begin{align}
\label{eq:loss_ssim}
\mathcal{L}_{ssim} =- SSIM(\mathbf{d}_m,\bar{\mathbf{d}}_m),
\end{align}
where $SSIM$ measures the structural similarity \cite{wang2004image} and $\bar{\mathbf{d}}_m$ is the depth map of the garment before deformation.

Finally, to prevent the garment from being obscured by other avatar components, we introduce a Visibility Loss $\mathcal{L}_{vis}$ which refines the positions of Gaussian points to ensure that all parts of the garment remain visible when it is combined with other inner layers of the avatar.
Intuitively, a garment should be closer to the camera than the covered human parts to remain visible. To achieve this, we optimize the depth value of the garment points $G_m$ to be lower than that of the corresponding avatar points in each camera view:
\begin{align}
\label{eq:loss_occ}
\begin{aligned}
\mathcal{L}_{vis} = max(0, -(\mathbf{d}_{av} - \mathbf{d}_o)*M_{m}+\delta _{occ})
\end{aligned}
\end{align}
where $\delta_{occ}$ is a margin gap set at 0.03.

\section{Experiments}
\begin{figure*}[t]
\centering
\includegraphics[width=1\linewidth]{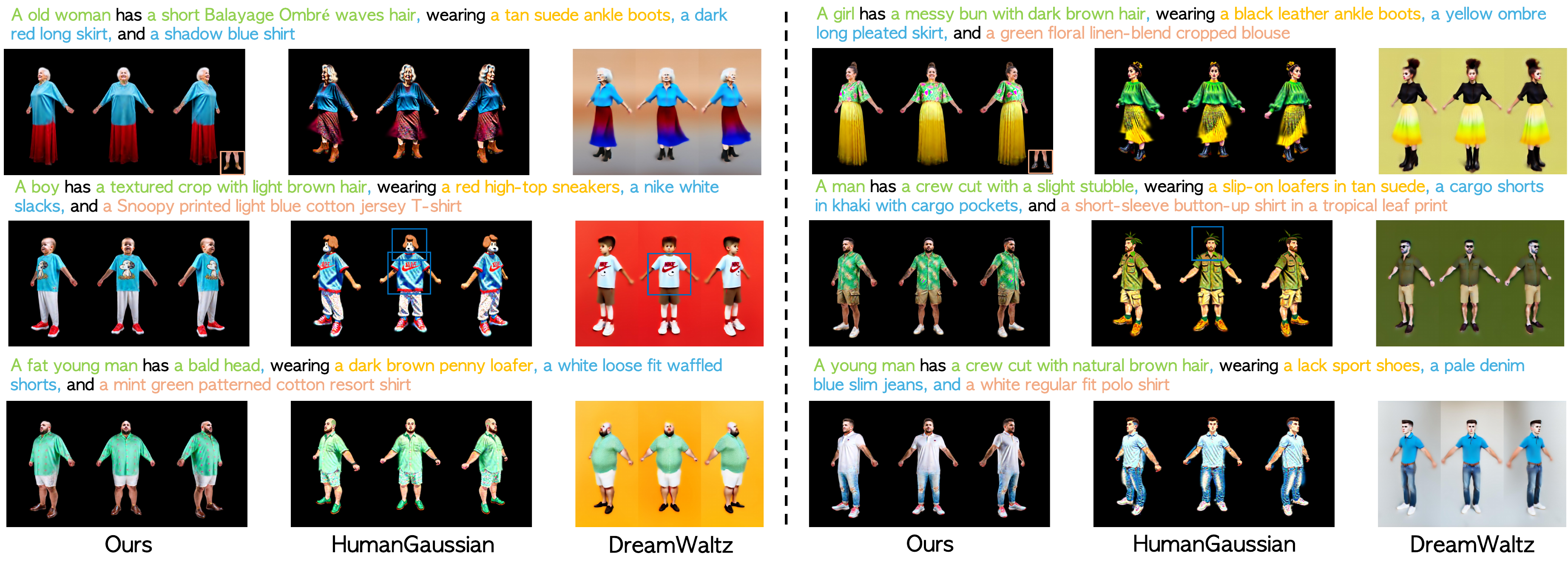}
\caption{Qualitative results. We compare our method with SOTA 3D human generators on six different prompts, each showing three camera views.
}
\label{fig:qualitative_results}
\end{figure*}

\begin{figure*}[t]
\centering
\includegraphics[width=1\linewidth]{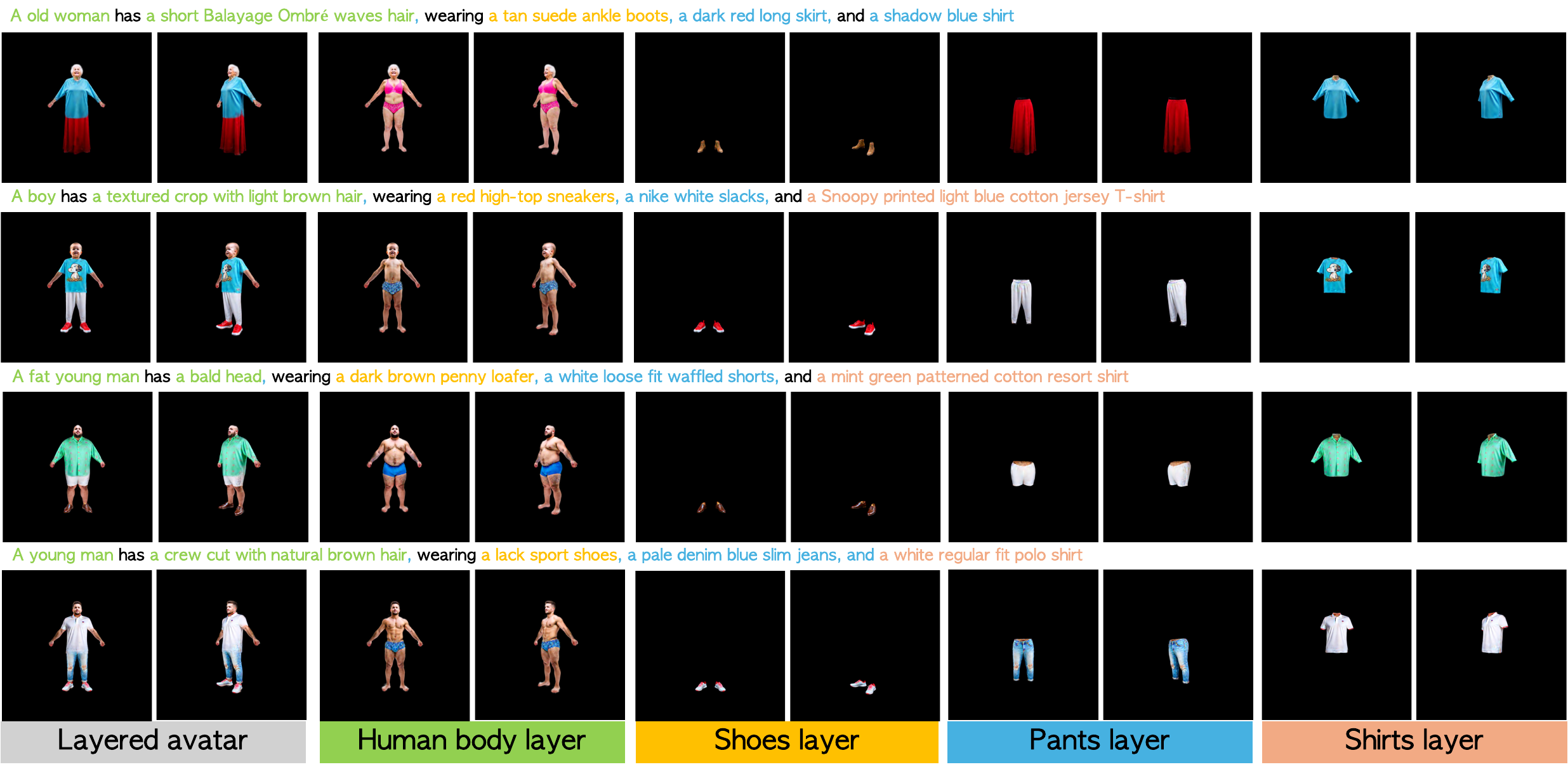}
\caption{Individual components for each avatar.}
\label{fig:decomposition}
\end{figure*}

\subsection{Implementation Details} We begin by sampling $5k$ points from SMPL to initialize sparse Gaussian points in each layer, subsequently densifying the Gaussian points four times to ensure high-quality avatar component generation. In each layer, we optimize the Gaussian points over $5k$ iterations with a batch size of 2, taking approximately 20 minutes on a single NVIDIA RTX 4090 GPU workstation. The samples generated by our models are rendered as images with a resolution of $1024 \times 1024$ for optimization purposes.
Given a text prompt in the format: "a \{human description\} has \{hair description\}, wearing \{garment description\}, \{garment description\}, ...", our method can automatically decompose the text description into multiple layer-specific text prompts and generate layers corresponding to each layer's text prompt for avatar modeling.

\subsection{Main Results}
\paragraph{Qualitative comparisons.}To evaluate the quality of the generated clothed human models, we compare our {\nickname} method with two state-of-the-art avatar generation models: DreamWaltz \cite{huang2024dreamwaltz} and HumanGaussian \cite{liu2023humangaussian}. The qualitative results are presented in Figure \ref{fig:qualitative_results}. 
As shown across the first row of Figure \ref{fig:qualitative_results}, the skirts generated by our approach exhibit more natural geometry as compared to existing methods. 
In the second and third rows of Figure \ref{fig:qualitative_results}, we also observe that the avatars generated by our method consistently align well with the given text prompts and capture more detailed features for each garment. 
Additionally, our avatars tend to look more photorealistic than those produced by HumanGaussian and contain more details {and finer textures} than those produced by DreamWaltz.
Overall, this qualitatively demonstrates our method's superior performance at rendering more realistic human appearances that are aligned with the text prompts, modeling more natural structures for both tight and loose garments, and capturing finer details for each avatar component.

\begin{table}
\centering
\caption{
User study: Ours vs HumanGaussian
}
\label{tab:user_study}
\begin{tabular}{@{}lc@{}}
\toprule
Comparison & Preference (\%) \\
\midrule
Texture quality &  \textbf{81.73} vs 18.27\\
Geometry quality &  \textbf{82.85} vs 17.15\\
Text Alignment &  \textbf{63.38} vs 36.62 \\
Reality & \textbf{93.85} vs 6.15\\
\bottomrule
\end{tabular}
\end{table}

\paragraph{Quantitative comparison.}  
We randomly selected 20 text prompts for avatar generation and compared our method with the state-of-the-art (SOTA) method, HumanGaussian \cite{liu2023humangaussian}. Specifically, we adapted the CLIP Score \cite{taited2023CLIPScore} to measure the alignment between the generated avatars and the given text, and used the Fréchet Inception Distance (FID) \cite{NIPS2017_8a1d6947} to evaluate the distribution gap between images rendered by avatars and a real 2D human dataset \cite{fu2022stylegan}. We find that our method consistently surpasses HumanGaussian on both metrics (e.g., \textbf{33.55} vs. 31.08 on CLIP Score ($\uparrow$) and \textbf{283} vs. 322 on FID ($\downarrow$)).

Moreover, we conducted a user study and followed \cite{liu2023humangaussian} to evaluate the quality of generated avatars from three aspects: (1) Texture Quality, (2) Geometry Quality, and (3) Text Alignment. Additionally, we {added a question on the realism aspect} to assess the photorealistic quality of avatars. As shown in Table \ref{tab:user_study}, our method consistently outperforms the SOTA across all the evaluated aspects.

\paragraph{Decomposition Ability.} As shown in Fig. \ref{fig:decomposition}, our avatars can be {conveniently} decomposed to a human body with a set of garments, where each avatar component {contains detailed appearance/textures} and high-quality geometry. 
{We note that this decomposability} further supports {users} to customize avatars easily.

\subsection{Ablation Study} \label{sec:cloth}

\begin{figure}
\includegraphics[width=0.45\textwidth]{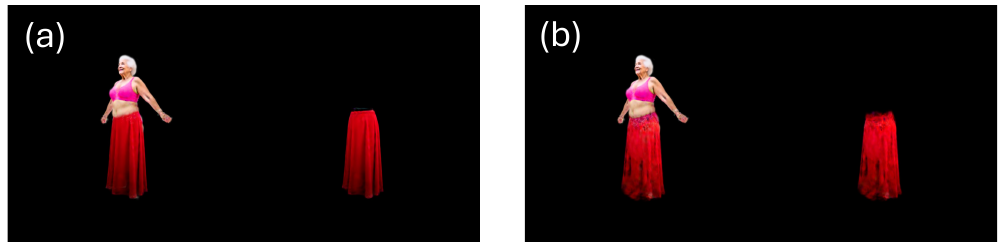}
\caption{Ablation for C2F. (a) Avatar w/ C2F. (B) Avatar w/o C2F.}
\label{fig:ab1}
\end{figure}
\paragraph{Impact of Coarse-to-Fine (C2F) strategy.}
As shown in Fig. \ref{fig:ab1}, directly optimizing Gaussian points sampled from SMPL without using our Coarse-to-Fine strategy may result in the generation of garments with geometric errors and large blurry areas.

\begin{figure}
\includegraphics[width=0.45\textwidth]{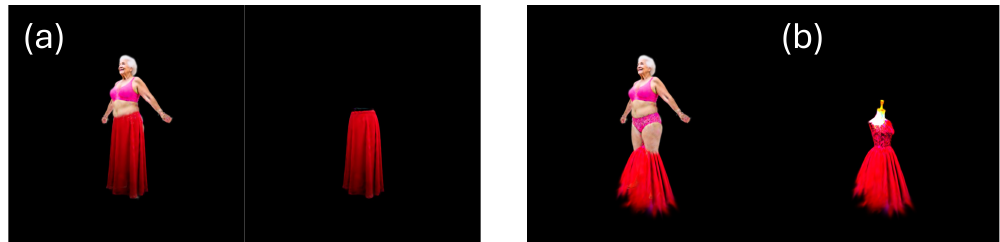}
\caption{Ablation for dual-SDS loss. (a) Avatar w/ dual-SDS. (B) Avatar w/o dual-SDS.}
\label{fig:ab2}
\end{figure}
\paragraph{Impact of dual-SDS loss.}
As shown in Fig. \ref{fig:ab2}, when replacing dual-SDS loss with a normal SDS loss, the generated garments tend to struggle to fit well with the human body.

\begin{figure}
\includegraphics[width=0.45\textwidth]{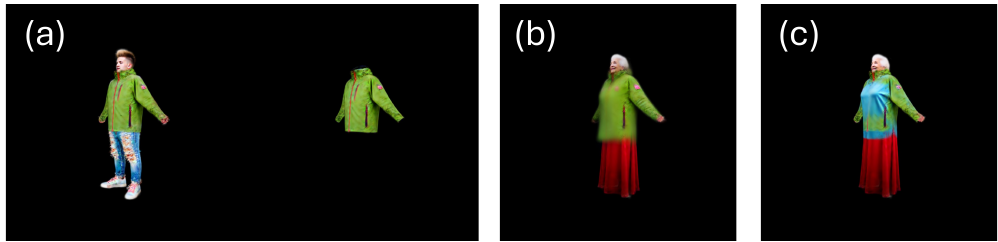}
\caption{Ablation for regularization loss. (a) Source Avatar. (B) Target avatar w/ regularization. (c) Target avatar w/o regularization.}
\label{fig:ab3}
\end{figure}
\paragraph{Impact of adaptation regularization loss.}
Directly transferring the garment from the source to the target avatar without regularization results in incoherence between the transferred garment and the target avatar (see Fig. \ref{fig:ab3}).
\section{Conclusion}

In this paper, we propose a LAGA, layered 3D human generation framework based on 3D GS, which generates decomposable clothed avatars with diverse garments and supports garment transfer across avatars with various shapes.
Our key insight lies in recognizing the high flexibility and controllability of GS, which unlocks significant potential for more flexible, layered avatar generation. Specifically, we introduce a coarse-to-fine generation strategy to facilitate diverse garment creation and a dual-SDS loss to ensure coherence between each avatar component. We also introduce three regularization losses to guide the movement of Gaussian points for garment adaptation. Extensive experiments demonstrate that our approach surpasses existing methods in generating 3D clothed humans.
{
    \small
    \bibliographystyle{ieeenat_fullname}
    \bibliography{sample-base}
}
\end{document}